\documentclass{cernyrep} 
\usepackage{texnames}
\usepackage[T1]{fontenc}
\begin{document}
\title{Production of hypertritons in heavy ion collisions around the threshold
of strangeness production}
 
\author{Christoph Hartnack$^1$, Arnaud Le Fèvre$^2$, Yvonne Leifels$^2$ and
J\"org Aichelin$^1$}

\institute{$^1$ SUBATECH, UMR 6457, Ecole des Mines de Nantes, IN2P3/CNRS,
Universit\'e de Nantes,\\ ~~~~~~ 4 rue Alfred Kastler, 44 307 Nantes, France \\
$^2$ GSI Helmholtzzentrum f\"ur Schwerionenforschung GmbH, Planckstra\ss{}e 1,  
64291 Darmstadt,\\ ~~~~~~  Germany}

\maketitle 

\begin{abstract}
We use the Isospin Quantum Molecular Dynamics approach supplemented with a
phase space coalescence to study the properties of the production of
hypertritons. We see strong influences of the hyperon rescattering on the
yields. The hypertritons show up to be quite aligned to the properties of
nuclear matter underlining the necessity of rescattering to transport the
hyperons to the spectator matter. 
\end{abstract}
 
\section{Introduction}
The production of hypernuclei as extension of the common periodic system or as
important ingredient for understanding strong interaction (see
e.g. \cite{schaffner,pano,hashimoto}) has recently gained strong interest for understanding the
properties of the hyperon interaction with nuclear matter, in particular since
the publication of recent results of experimental collaborations 
\cite{star,saito,alice,rap13}.
Several theoretical approaches have proposed the combination of transport and 
fragmentation models in order to understand the data
\cite{gaitanos,steinheimer,botvina}.
Very recently a novel fragmentation approach, FRIGA \cite{arnaud}, based 
on the maximization of the binding energy of the fragments, succeeded in 
explaining FOPI data on hypertritons as well as brandnew data of HypHI 
\cite{rap15} concerning hypertritons and $^4\!\! _\Lambda$H.

\begin{figure}[bt]
\centering\includegraphics[width=.45\linewidth]{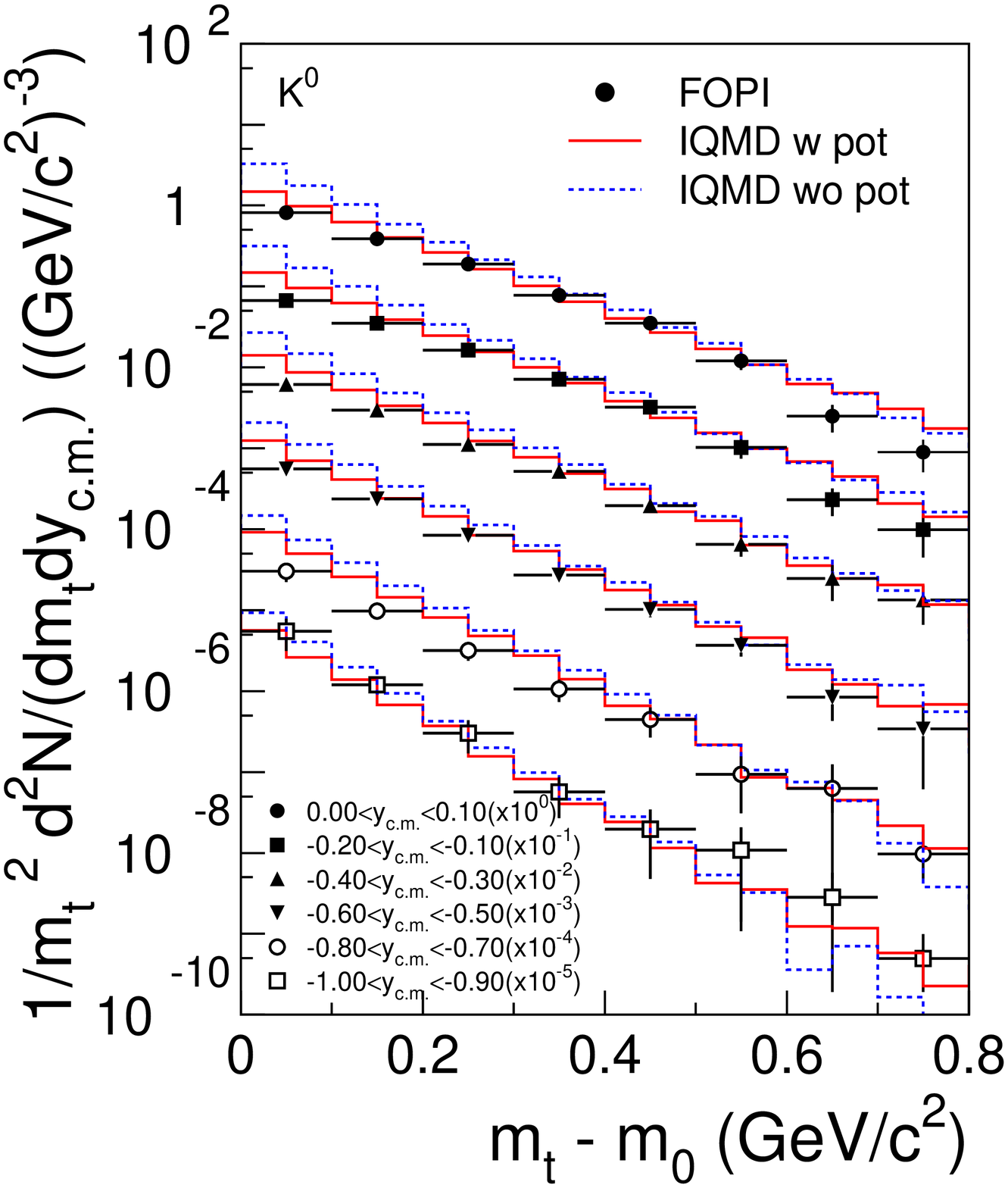}\includegraphics[width=.45\linewidth]{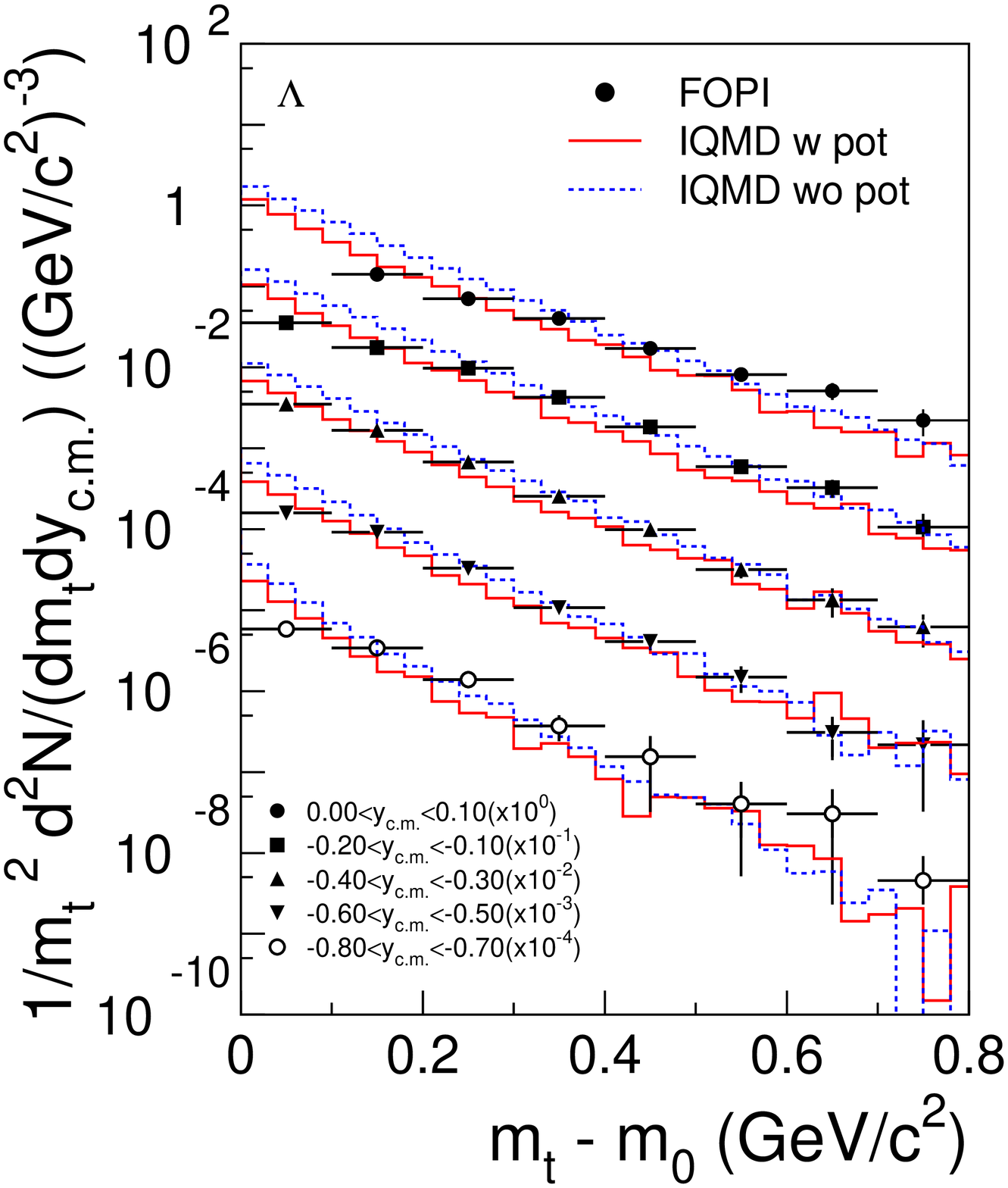}
\caption{Comparison of $K^0$ and hyperon spectra of Ni+Ni collisions between FOPI
data and IQMD calculations }
\label{fig:spectra}
\end{figure}

In this contribution we use a transport model, IQMD \cite{IQMD}, supplemented by a phase
space coalescence to study the properties of hypertritons in
collision of Ni+Ni at an incident energy of 1.93 AGeV. This model shows quite
comparable results for the hypertriton yields as the much more sophisticated 
analysis presented in \cite{arnaud}.

\section{Production of hyperons in heavy ion collisions}
In this section we want only to sketch the production of hyperons in
nuclear matter and refer for a detailed discussion to \cite{Hartnack-PR}.
At energies around the threshold kaons are dominantly produced in multistep
processes using resonances (especially the $\Delta$) as intermediate energy
storage. The energetically most favorable channel is the production of a kaon
together with a hyperon, this channel being also the major channel for the hyperon
production. This common production channel links the effects acting on thekaon
production to that of the hyperon. Thus the hyperon yields are also depending on the
nuclear equation of state and on the kaon optical potential: a soft EOS reaches
higher densities than a hard one and thus causes a smaller mean free path of the nucleons which
enhances the collision rate. This allows more deltas to undergo a second high
energetic collision (and thus to produce a kaon-hyperon pair) before decaying
again into a nucleon and a pion. However at these high densities the kaon
optical potential causes a penalty on the strangeness production by enhancing
the thresholds and thus lowering the production cross section for a given
energy. This causes a reduction of the production of kaon-hyperon pairs with
respect to calculations with a kaon optical potential.

This effect can be seen in Fig.~\ref{fig:spectra} where we compare the spectra of
$K^0$ and $\Lambda$ measured by FOPI in collisions of Ni+Ni at 1.93 AGeV
incident energy and IQMD calculations performed with (full line) and without (dotted
line) kaon optical potential.
We see that the potential penalizes as well kaons as hyperons. The spectra are
in good agrement, however it should be noted that the calculated temperatures for
hyperons are somehow lower than the experimental values. It should also be
noted that the rapidity distributions of the kaons measured by FOPI and KaoS
can be well described.

\begin{figure}[hbt]
\centering\includegraphics[width=.45\linewidth]{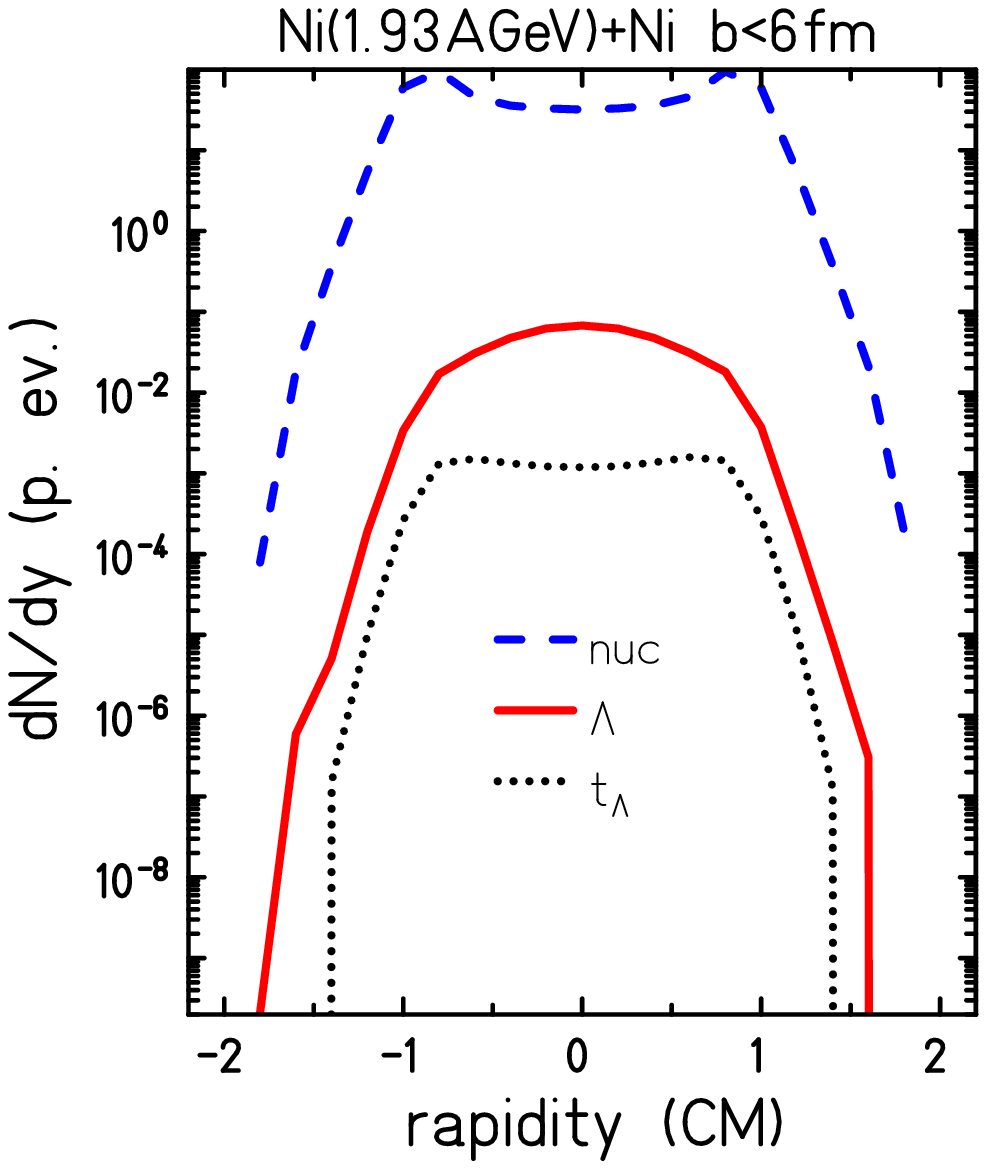}\includegraphics[width=.45\linewidth]{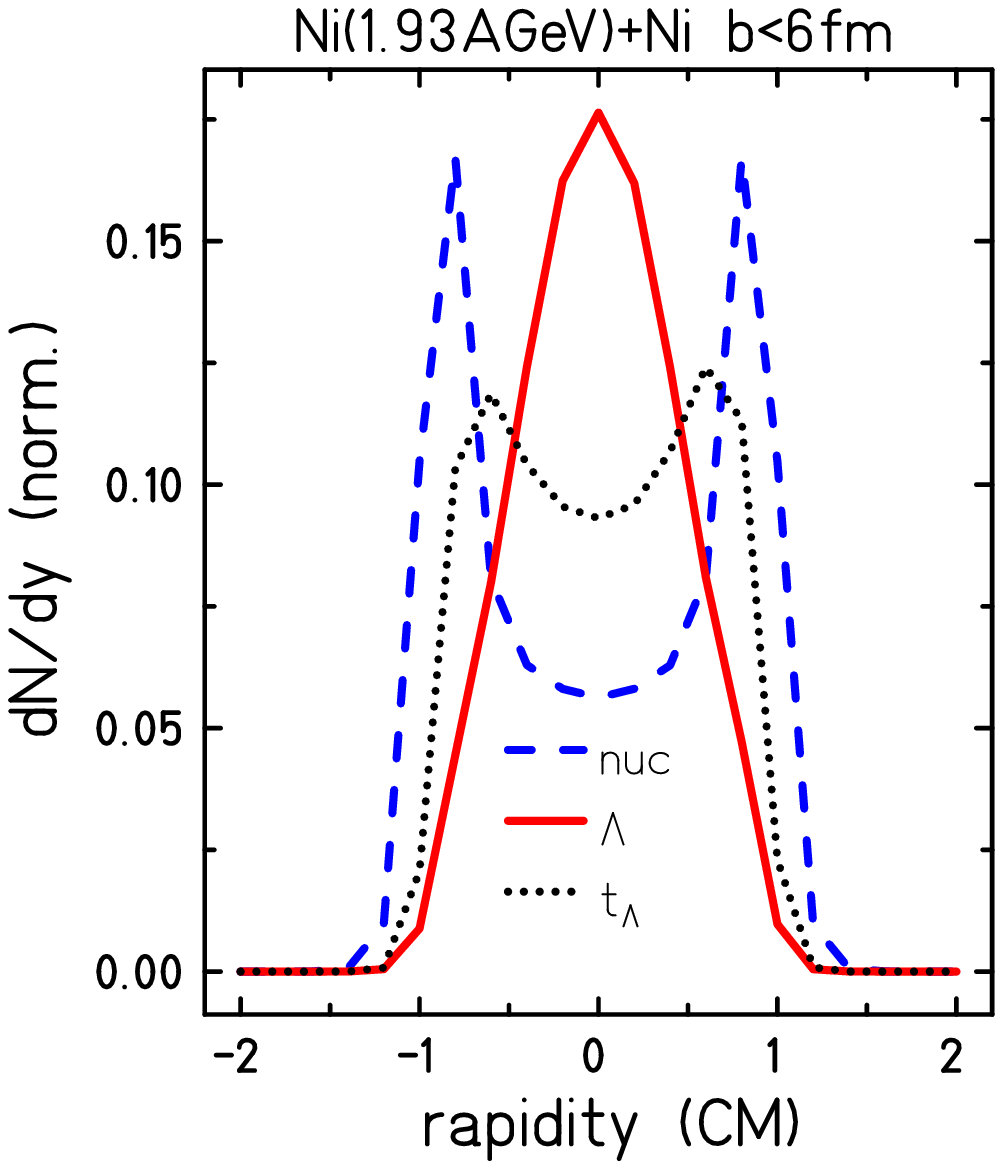}
\caption{Rapidity distributions of nucleons,hyperons and hypertritons in an
absolute yield (left) and normalised to the total yield.}
\label{fig:rapidity}
\end{figure}

Let us now compare the rapidity distributions of nucleons, hyperons and
hypertritons. Fig.\ref{fig:rapidity} shows on the l.h.s the absolute yields as
function of the rapidity. The ordinate is presented in logarithmic scale due to
the large differences in absolute yields. This nicely demonstrates that the
production of hypertritons is really a rare event.
The r.h.s. shows the same distribution in linear scale but normalized to the
particle yields. We see clearly that nucleons and hyperons are peaked in
completely different regions in phase space. While the nucleon (dashed lines)
peak around projectile and target rapidity, which means that the nuclear matter is not at
all stopped, the hyperons (full line) peak around midrapidity. This is due to the effect
that their production points are distributed around the cm of the colliding
system which is typically the centre-of-mass of a NN system. It should be noted, 
that the production follows the kinematics of a 3 body phase space decay, where the hyperon has
the highest mass - and thus the lowest velocity - of the 3 outgoing particles. Thus
the hyperons show relatively low momenta at production in the NN centre of mass.   
The hypertritons (dotted line) have to combine Lambdas with nuclear spectator matter: their
production peak lies in between the distributions of nucleons and hyperons but
with a strong dominance of the nuclear matter.

\section{The role of hyperon rescattering}
\begin{figure}[hbt]
\centering\includegraphics[width=.45\linewidth]{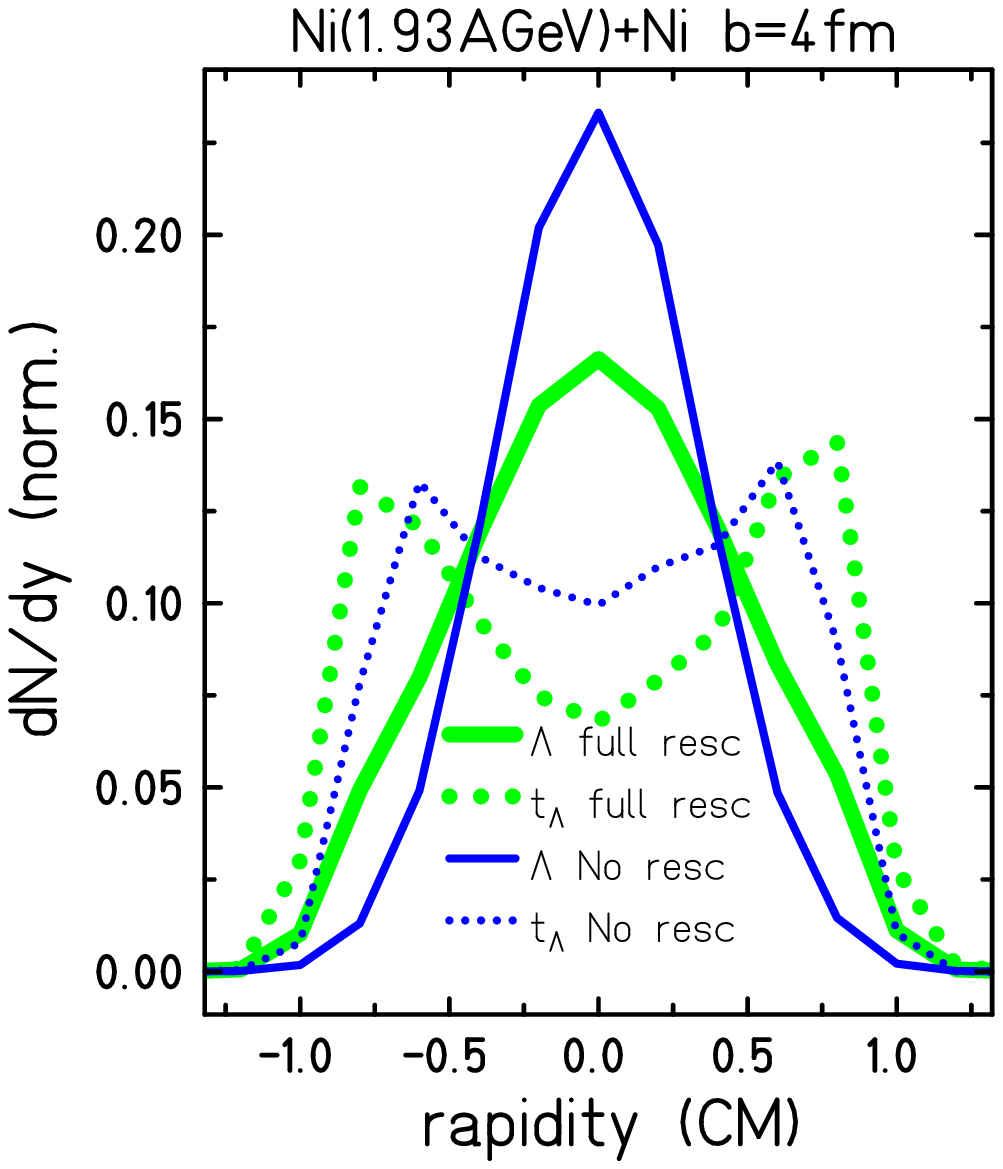}
\includegraphics[width=.45\linewidth]{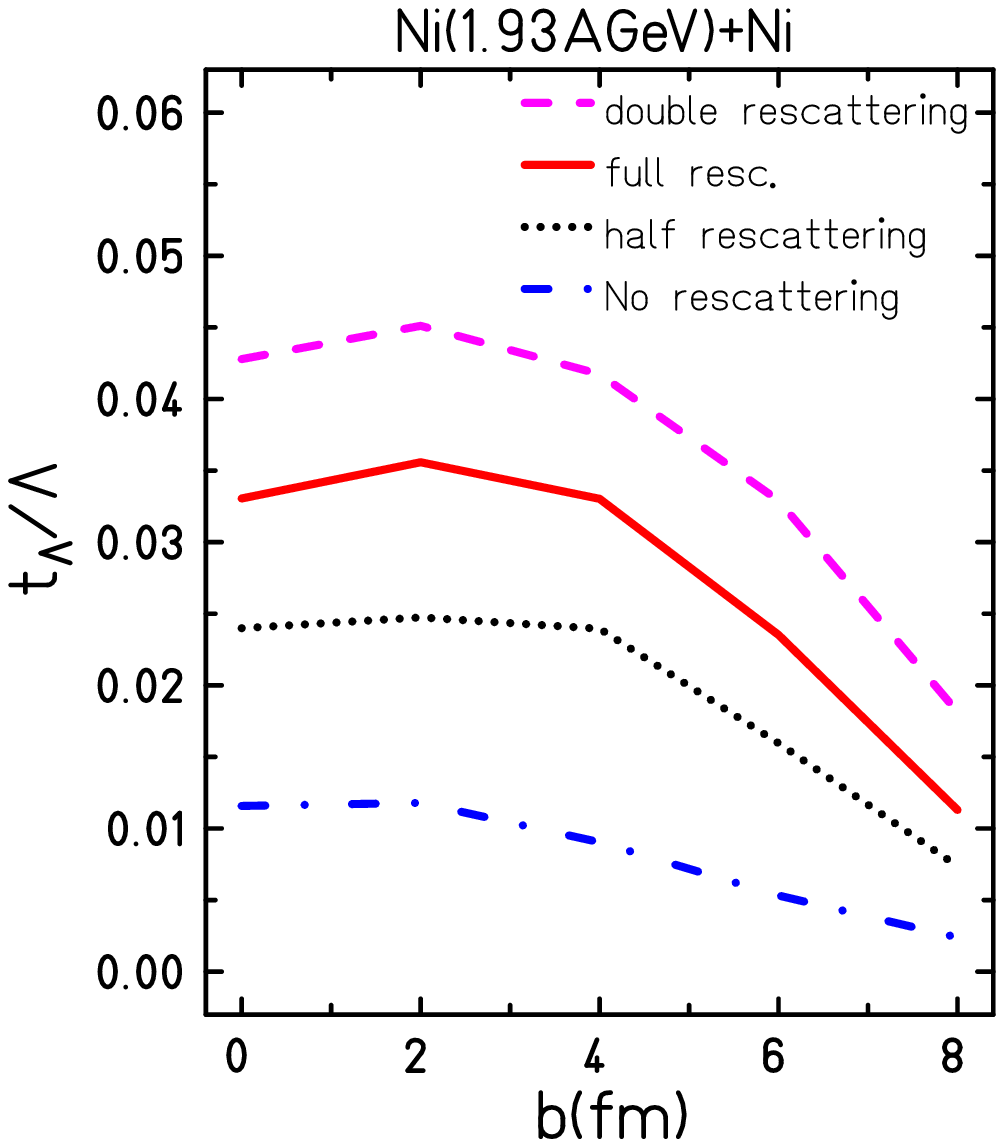}
\caption{Normalised rapidity distributions of hyperons and hypertritons with
and without rescattering (left) and the ratio hypertritons/hyperons as a
function of the impact parameter for different assumptions on the rescattering
cross sections (right)}
\label{fig:resc}
\end{figure}
The final rapidity distribution of the hyperons reflects not only the kinematics
of the production process but also the rescattering with nuclear matter as it can be
depicted from the l.h.s. of fig \ref{fig:resc}, where we plot the normalized
rapidity distributions of hyperons (full lines) and hypertritons (dotted lines).
The rapidity distribution of hyperons with rescattering disabled (blue lines)
correspond to the distribution at production and is therefore quite narrow.
It is the rescattering (green thick lines) that enhances the momenta of the hyperons and
thus leads to higher temperatures and broader rapidity distributions.
This also influences the rapidity distributions of the hypertritons which need
as well hyperons and nucleons for being created. The rapidity distribution of
hypertritons is thus also broader if rescattering is allowed.

Besides the shape of the rapidity disctributions of hypertritons, the
rescattering influences also the yield of hypertritons significantly. This can
be seen on the r.h.s. where the ratio hypertriton/hyperon is shown as function of
the impact parameters using different options for the rescattering.
The calculations using full rescattering (full lines) yield nearly 3 times more 
hypertritons than calculations disabeling the rescattering (dash-dotted
lines). If we set the rescattering cross sections to the half of its value
(dotted lines) we still obtain nearly the double yield as in the calculations without
rescattering. This can be easily understood from the rapidity distribution on
the l.h.s. of fig \ref{fig:resc}: without rescattering the distribution of hyperons becomes this kind of
narrow that only few hyperons can reach the region of spectator matter. However
it is that region which is fertile for the production of fragments since it is
there where clusters may remain undestroyed.

We can thus conclude that the main processus for creating hypertritons is to
transport hyperons via rescattering to the region of the spectator matter where
it may insert into a nucleon cluster. When having joined a cluster the
hyperon-nucleon potentials help to keep the fragment stable. In this context we
want to indicate that in IQMD the nucleons propagate by 2 and 3 body
interactions of Skyrme, Yukawa, and Coulomb type, supplemented by momentum
dependent interactions and asymetry potentials. Hyperons only interact by Skyrme
interactions assuming a factor of two third in the strength of the potential.
For details see \cite{Hartnack-PR}. 

The ratio of hypertritons/hyperons has the advantage to compensate other effects
on the hypertriton yield steming from the absolute hyperon yield. 
Effects acting on the kaon numbers like the equation of state or kaon optical 
potentials of course influence the absolute yield of hypertritons.

\begin{figure}[hbt]
\centering\includegraphics[width=.45\linewidth]{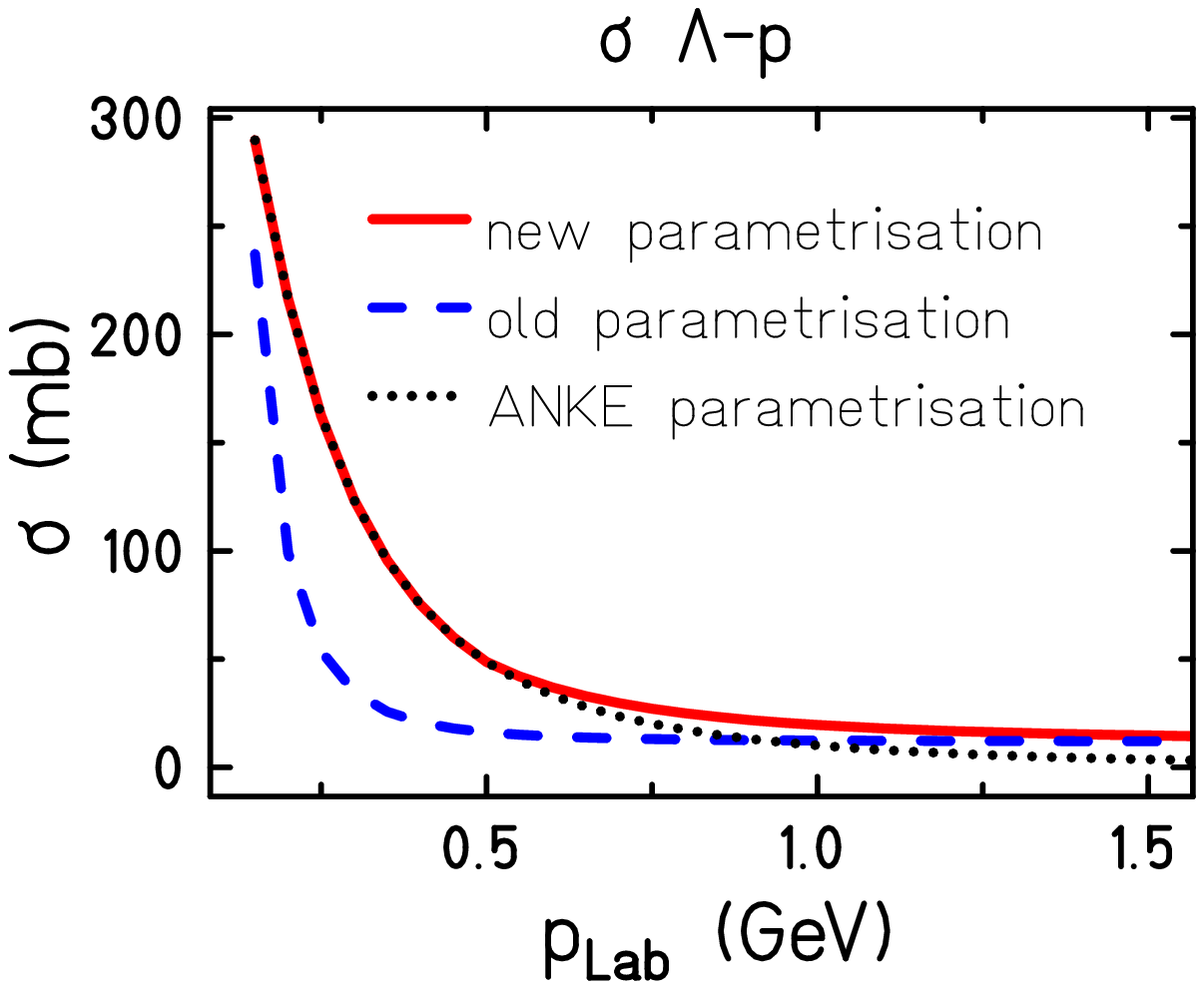}
\includegraphics[width=.3\linewidth]{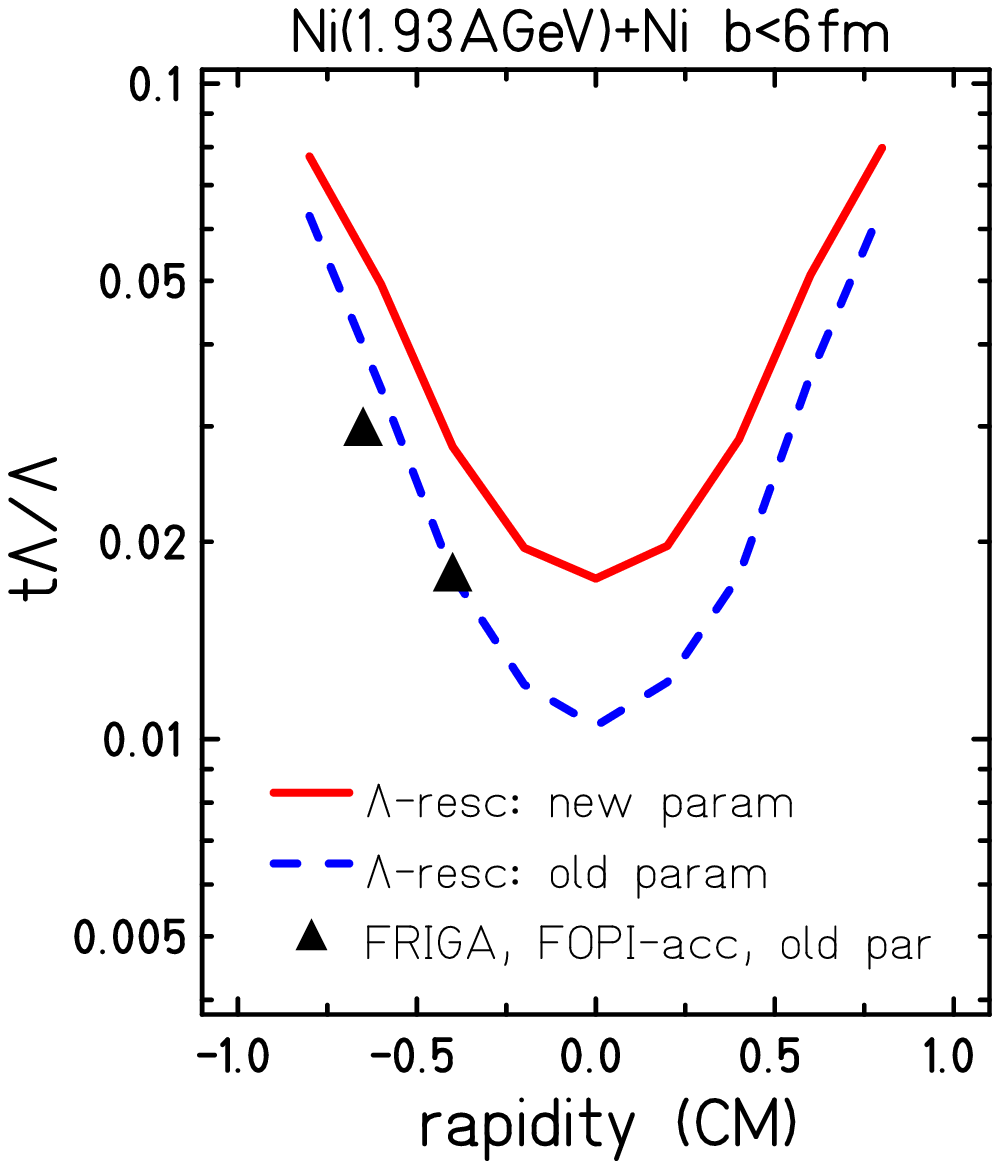}
\caption{Description of two parametrizations of rescattering and the hypertriton
to hyperon ratio as function of rapidity.}
\label{fig:ratio}
\end{figure}
Taking into account that the calculated hyperon spectra get lower temperatures 
than the experimental values we investigated the use of
a novel cross section parametrisation inspired by ANKE data \cite{ANKE}. 
Fig.\ref{fig:ratio} presents on the l.h.s. the ``old'' parametrisation (dashed line) and the new
fit (full line) and on the r.h.s. its effect on the hypertriton to hyperon ratios
as fucntion of rapidity. We see that the enhanced cross sections also raise the
ratios. For comparison we included results of a more sophisticated analysis
\cite{arnaud} but which was still using the old parametrisation. That
sophisticated analysis uses a minimum binding energies approach and applies the
acceptance cuts of FOPI. We see a good agreement of our simplified model with
these calculations which conforts us in using that simplified model for analysing 
the properties of hypertritons in detail.

\section{Properties of hypernuclei}
\begin{figure}[hbt]
\centering\hbox{\includegraphics[width=.35\linewidth]{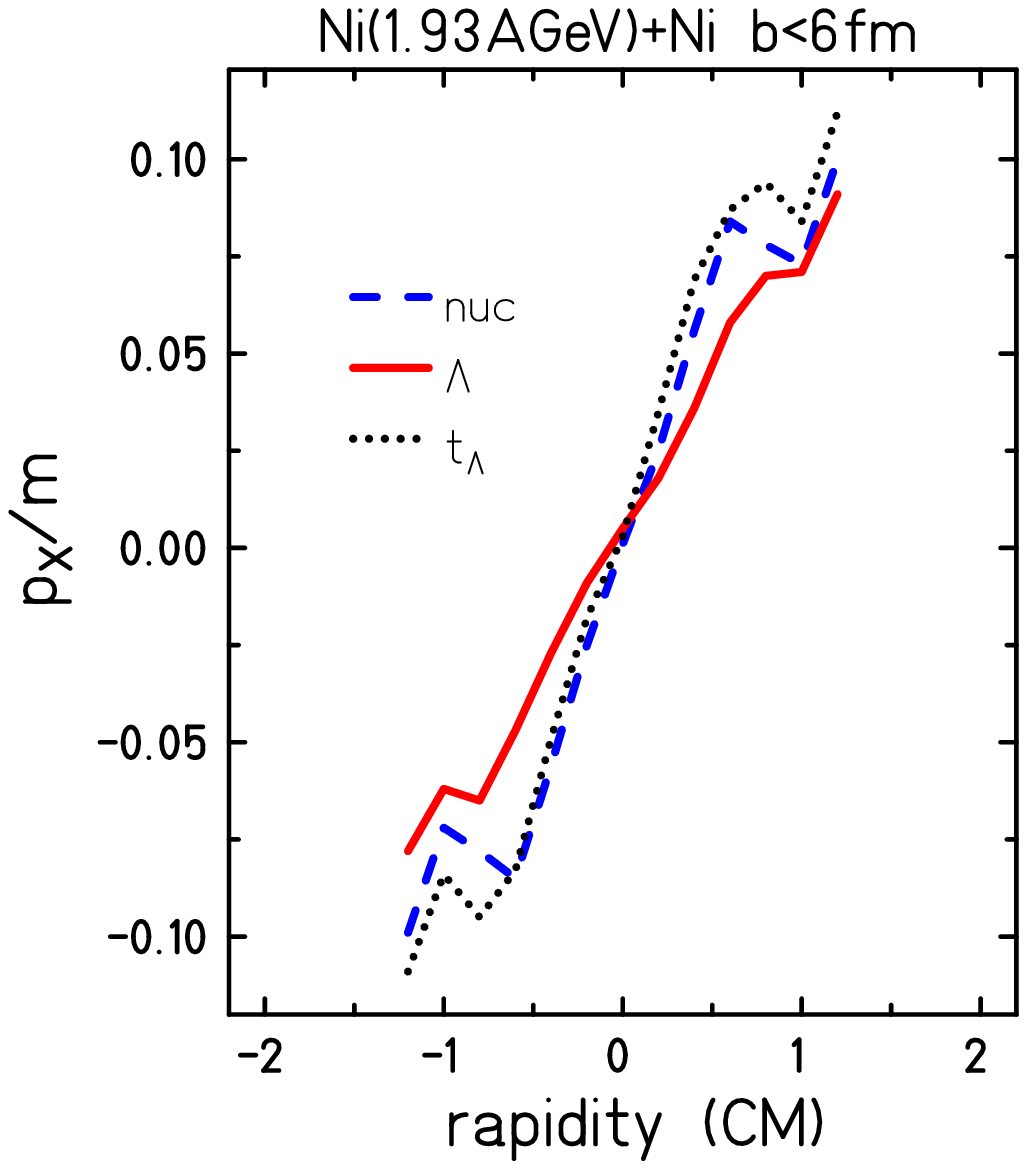}
\includegraphics[width=.32\linewidth]{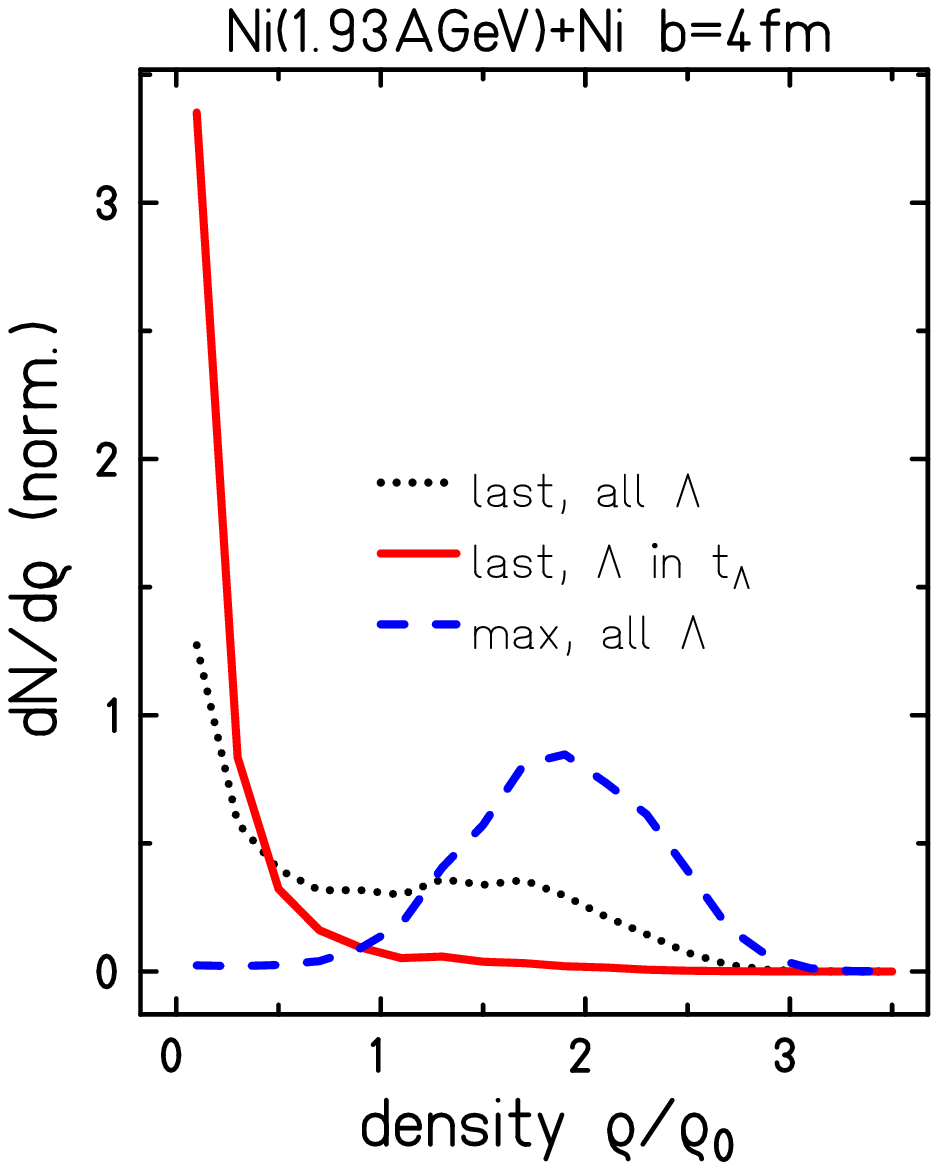}
\includegraphics[width=.29\linewidth]{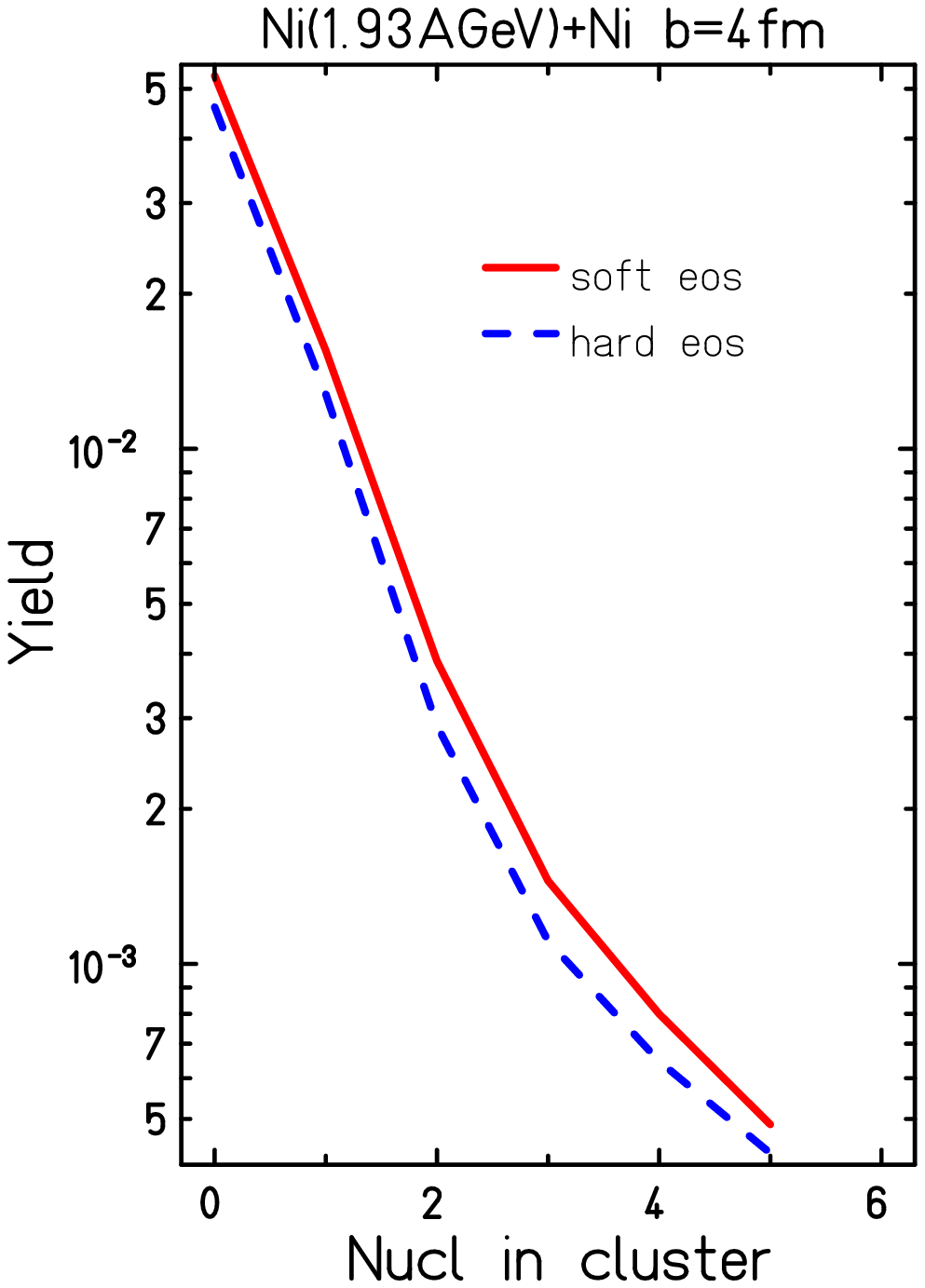}}
\caption{Left: Transverse flow divided by the mass as function of the rapidity
for Lambdas, nucleons and hypertritons.
Mid: Maximum densities experienced by the Lambdas and densities at last contact
for all Lamdas and Lamdas in hypertritons.
Right: Yield of the clusters as a function of the numbers of nucleons in
the cluster.}
\label{fig:pxrho}
\end{figure}
In order to show the correlation of hypertritons to nuclear matter let 
us compare the transverse flow of Lambdas (full line), nucleons
(dashed line) and hypertritons (dotted line), shown on the l.h.s. of 
Fig.~\ref{fig:pxrho}. We see that Lambdas show already
a significant flow, which is dominantly due to the rescattering of the hyperons
with the nuclear matter. That nuclear matter itself shows an even higher flow.
Hypertritons show a flow compatible with the rest of the nuclear matter, which
underlines the alignment of hypernuclei to the nuclear matter. It is due to a
large number of collisions that the hyperons enters into the cluster.

This mechanism is supported by the observation of the freeze-out densities  
in the mid of Fig.~\ref{fig:pxrho}: while the maximum density  
that a hyperon experienced (dashed line), typically the density of its
production, 
is around twice time nuclear density, the freeze-out density (dotted line), 
i.e. the density  of the last collisional contact, is quite lower.
This indicates that the collisions persist up to a late phase of the expansion
of  the nuclear matter. If we regard the hyperons bound in a hypertriton (full
line), they pratically all freezed out at densities well below normal matter
density. 

Let us now look on the properties of hypernuclei at different cluster size. 
In the following we will decribe the hypernuclei by the number of accompanying nucleons in the cluster.
Zero means a single unclustered hyperon and serves to underline the difference between unclustered hyperons and
hyperons in a cluster. As already indicated, the Lambda has to join the region
of spectator matter in order to integrate a cluster.
Since these regions are quite far away from the
distributions of hyperons (see fig \ref{fig:rapidity}), their production is of
course extremely suppressed. This finding is confirmed on the r.h.s. of fig
\ref{fig:pxrho}, which describes the yield of the clusters as a function of the
cluster size. The slight dependence of hypercluster yield on the nuclear
equation of state is due to the effect of the EOS on the hyperon production.
As already mentionned a soft EOS (full line) yields a higher hyperon yield than
a hard EOS (dashed line) and thus more hyperclusters can be formed.

\begin{figure}[hbt]
\centering\includegraphics[width=.3\linewidth]{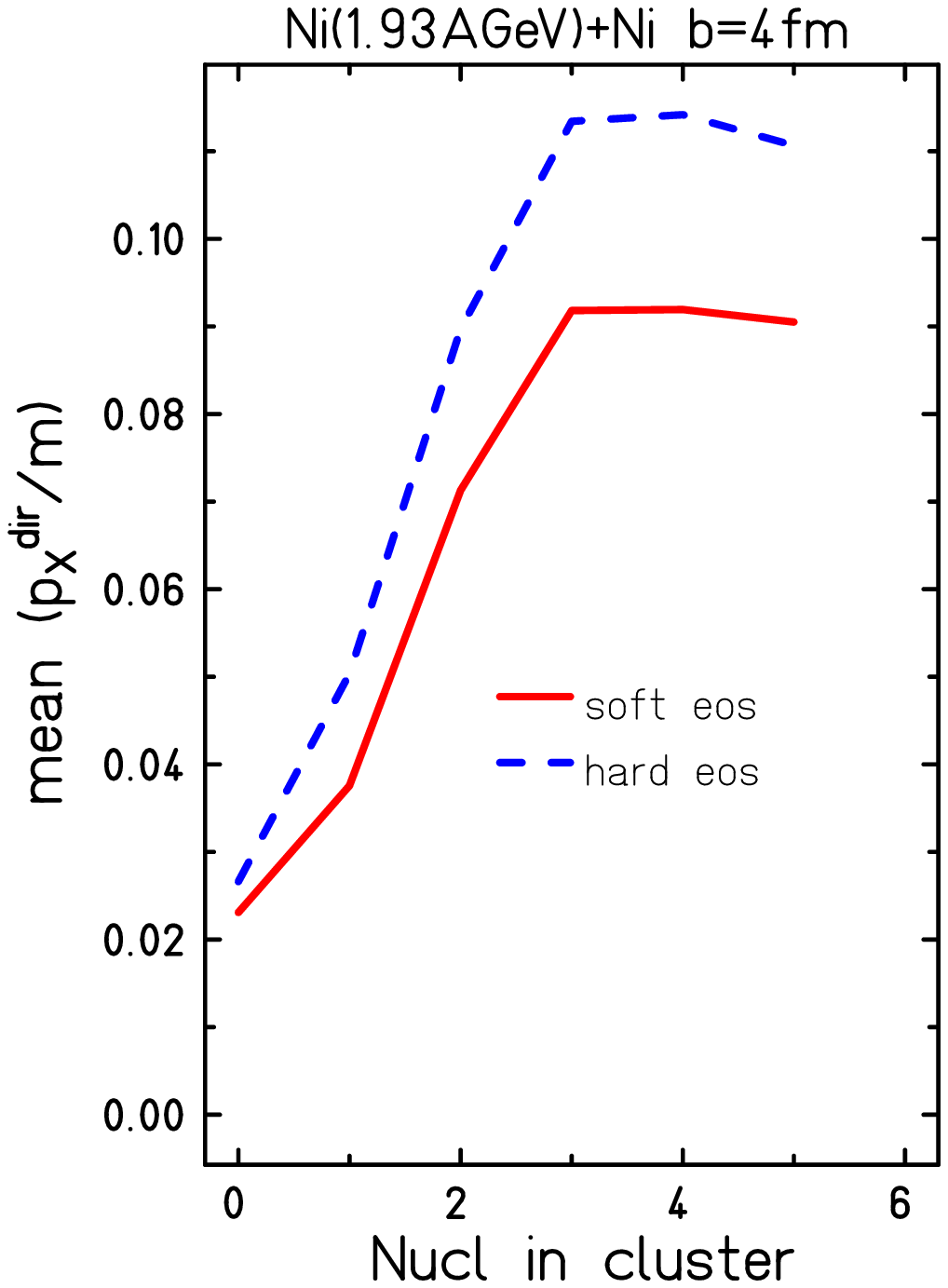}
\includegraphics[width=.3\linewidth]{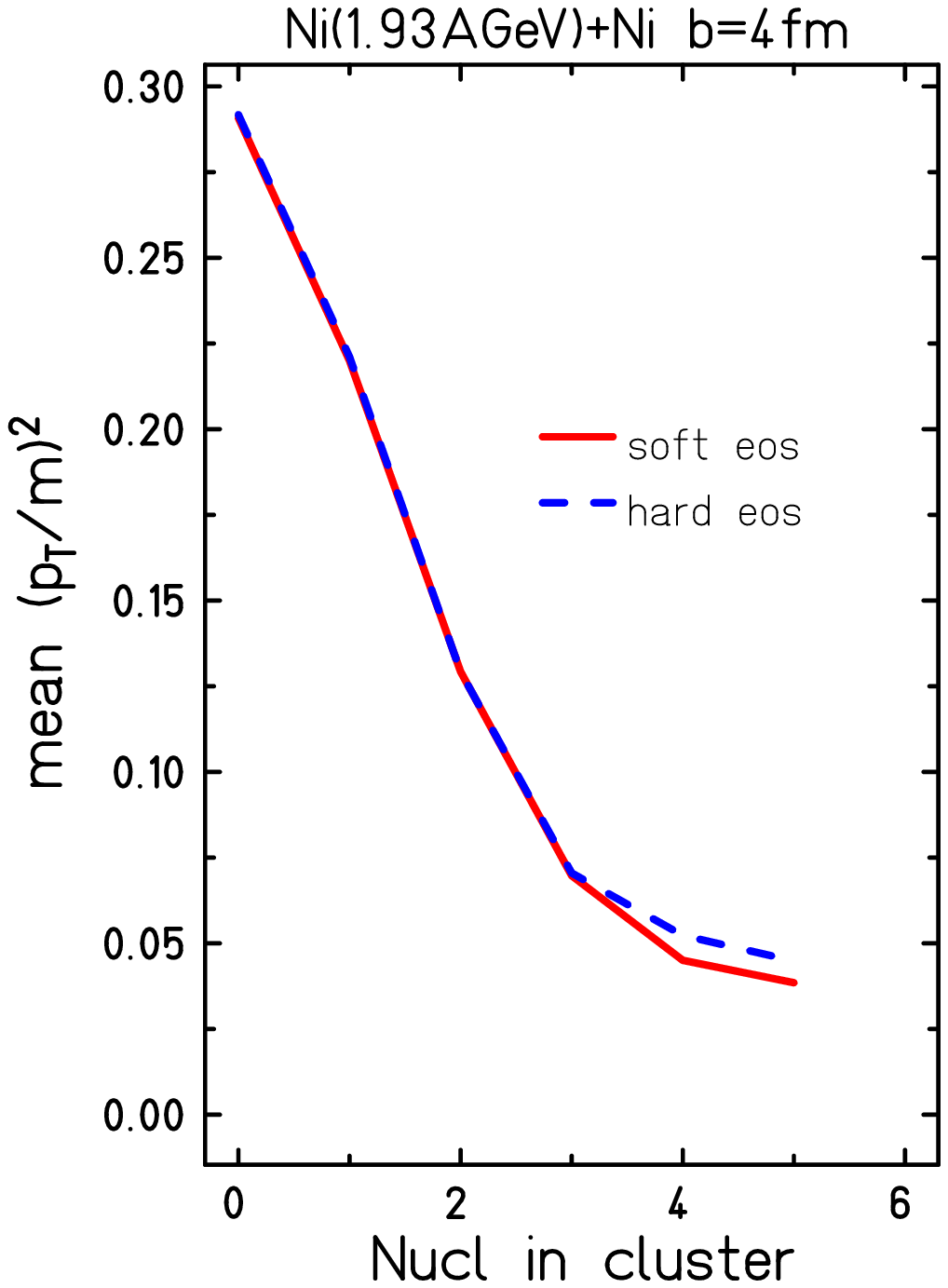}
\includegraphics[width=.3\linewidth]{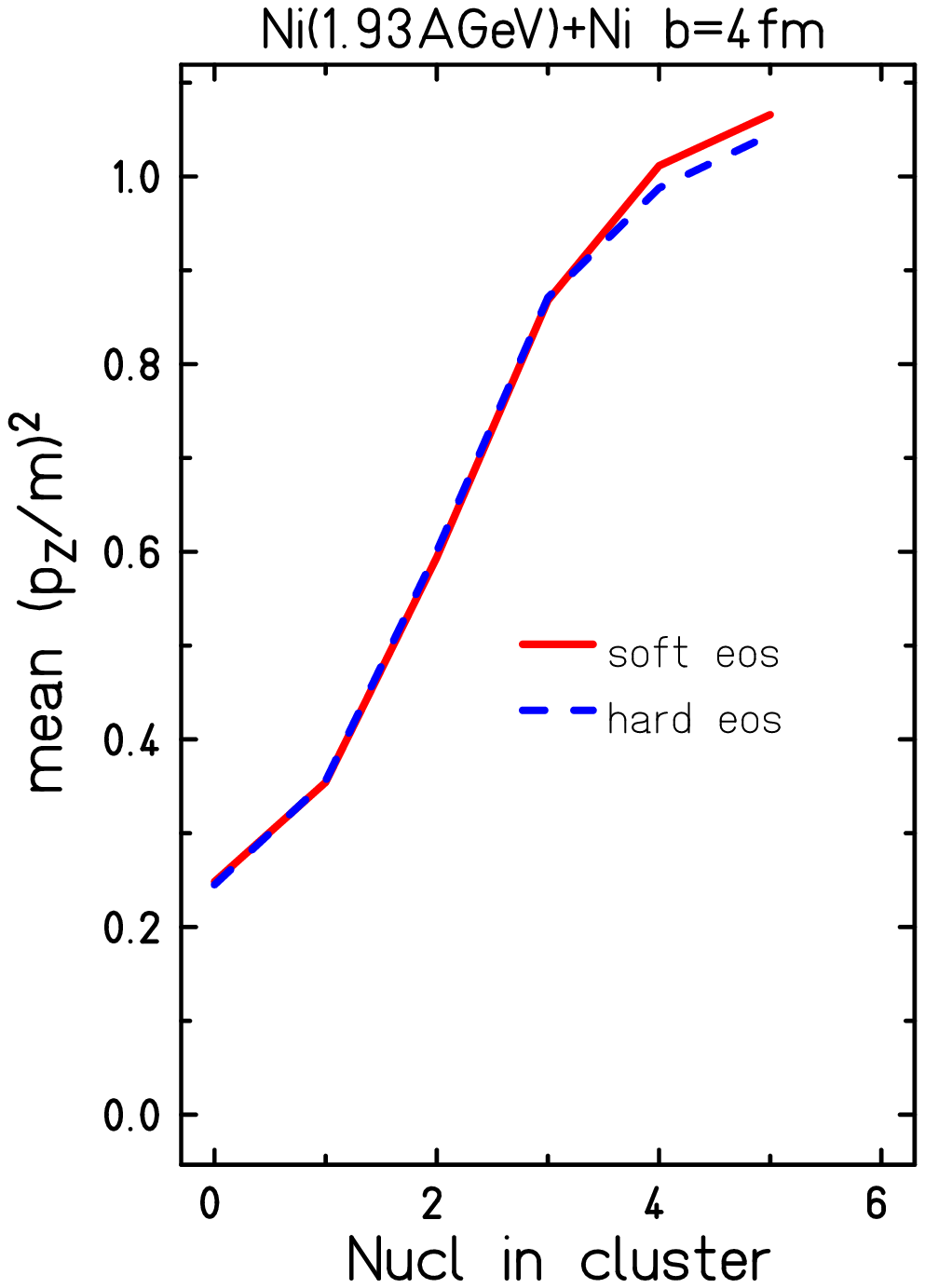}
\caption{Directed flow, mean quadratic transverse momentum and mean transverse longitudinal
momentum of the hyperon as function of the cluster size}
\label{fig:flow}
\end{figure}
Let us first look on the dynamical observables of the clusters. Fig
\ref{fig:flow} shows on the l.h.s. the directed flow (normalised to the total mass)
of the hyperclusters. As already seen on the l.h.s. of Fig.~\ref{fig:pxrho} even 
unclustered hyperons show a directed flow due to recattering.  Clustered hyperons 
show an enhanced flow, which
increases with fragment size, an effect which is already known from the
behaviour of normal nuclear fragments (``The fragments go with the flow'').
At higher cluster sizes the flow value seems to saturate. It should also be
noted that this observable shows a dependence on the nuclear equation of state,
similar to the behaviour known for normal nuclear matter: a hard equation of
state causes a higher directed flow.

The mid part and r.h.s. of  Fig \ref{fig:flow} show respectively the squared
transverse and longitudenal momenta, again normalised by the mass. 
The transverse momentum decreases with fragment size while the longitudinal
momentum increases.This reflects the effect that large clusters can only be
found at the projectile/target remnants which remain practically at
projectile/target rapidities. This supports the previous statement that the
production of large hyperclusters is stronly suppressed by the effect that only
few hyperons enter the region of spectator matter.

\section{Freeze-out of hypernuclei}
\begin{figure}[hbt]
\centering\includegraphics[width=.3\linewidth]{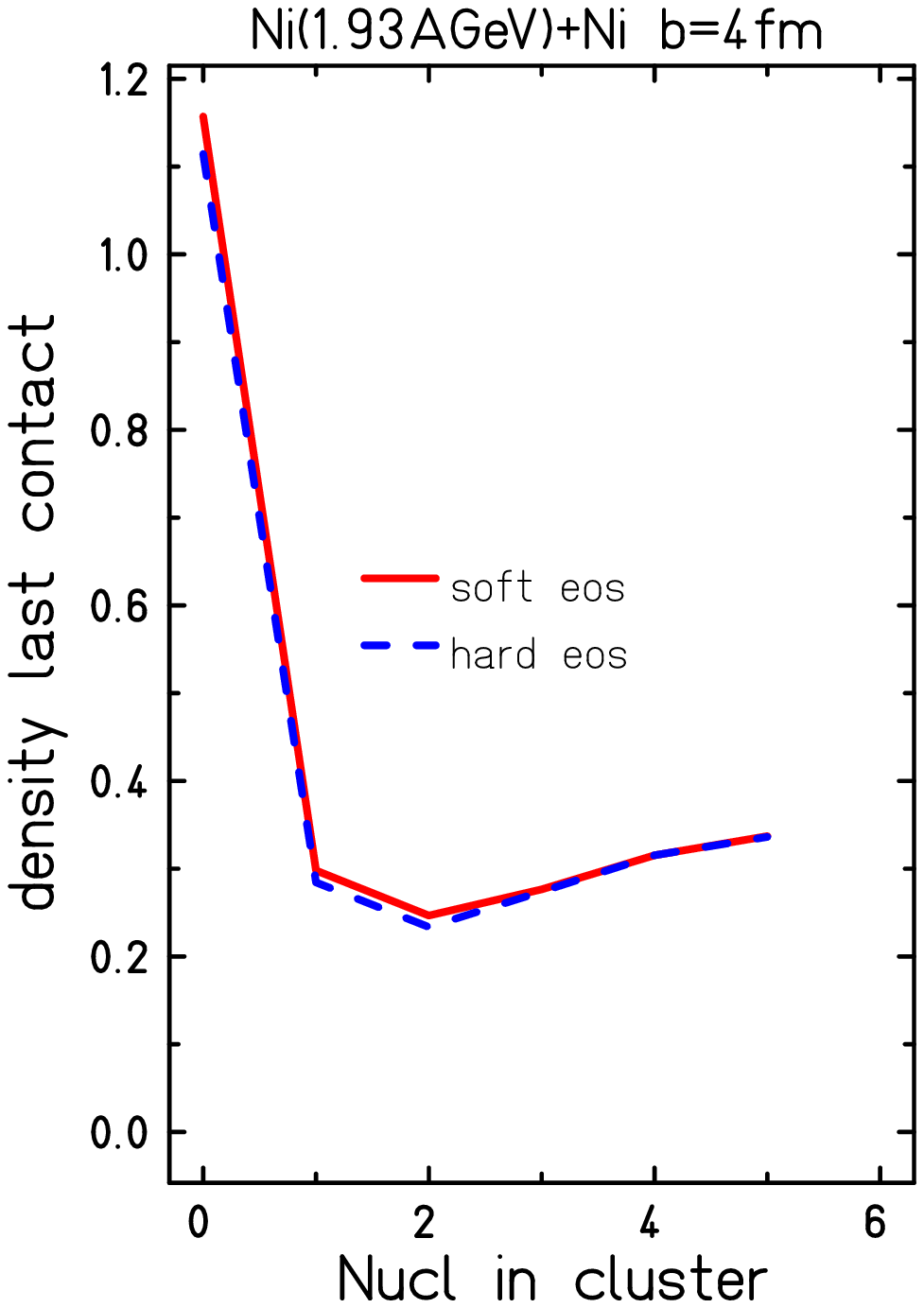}
\includegraphics[width=.3\linewidth]{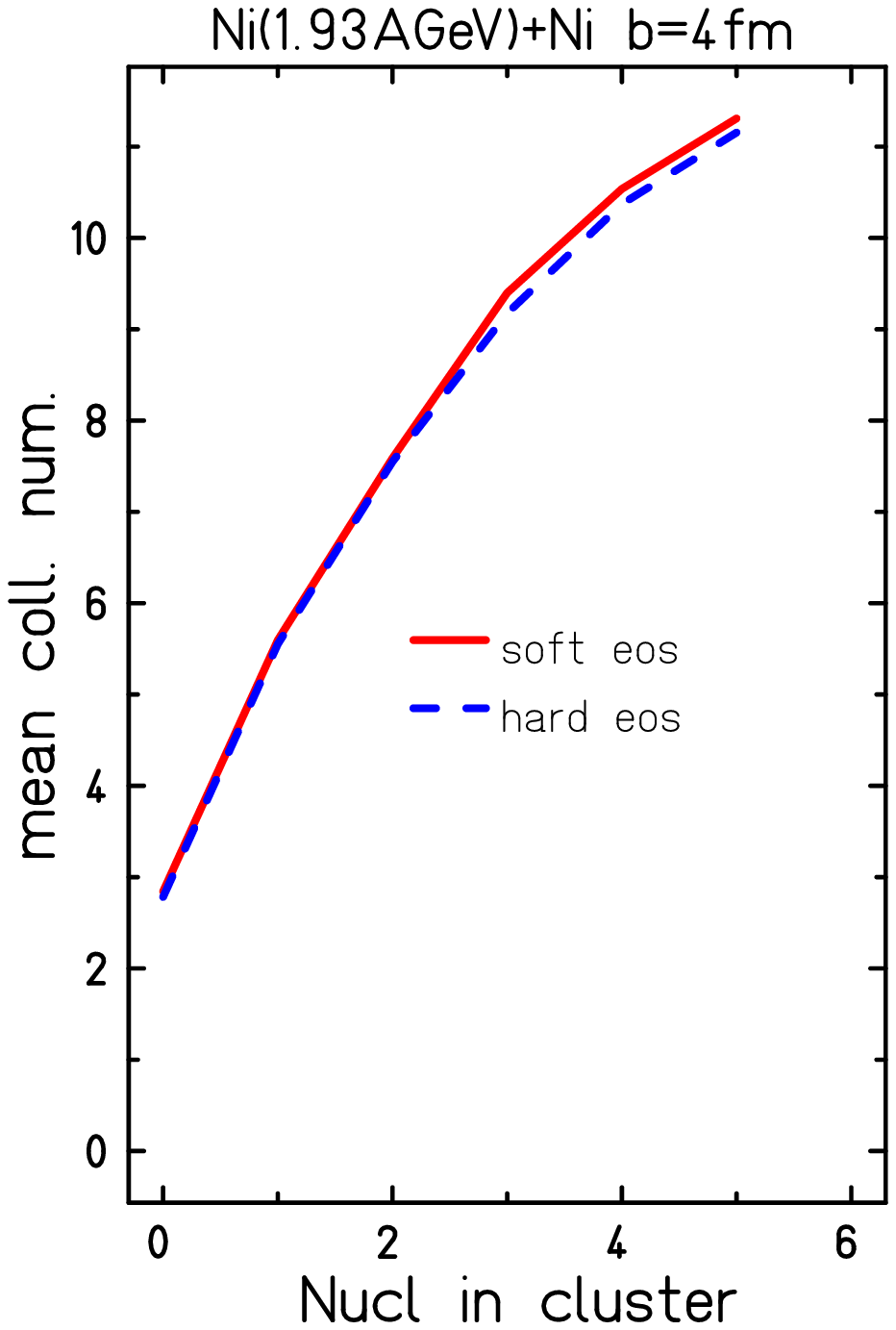}
\includegraphics[width=.3\linewidth]{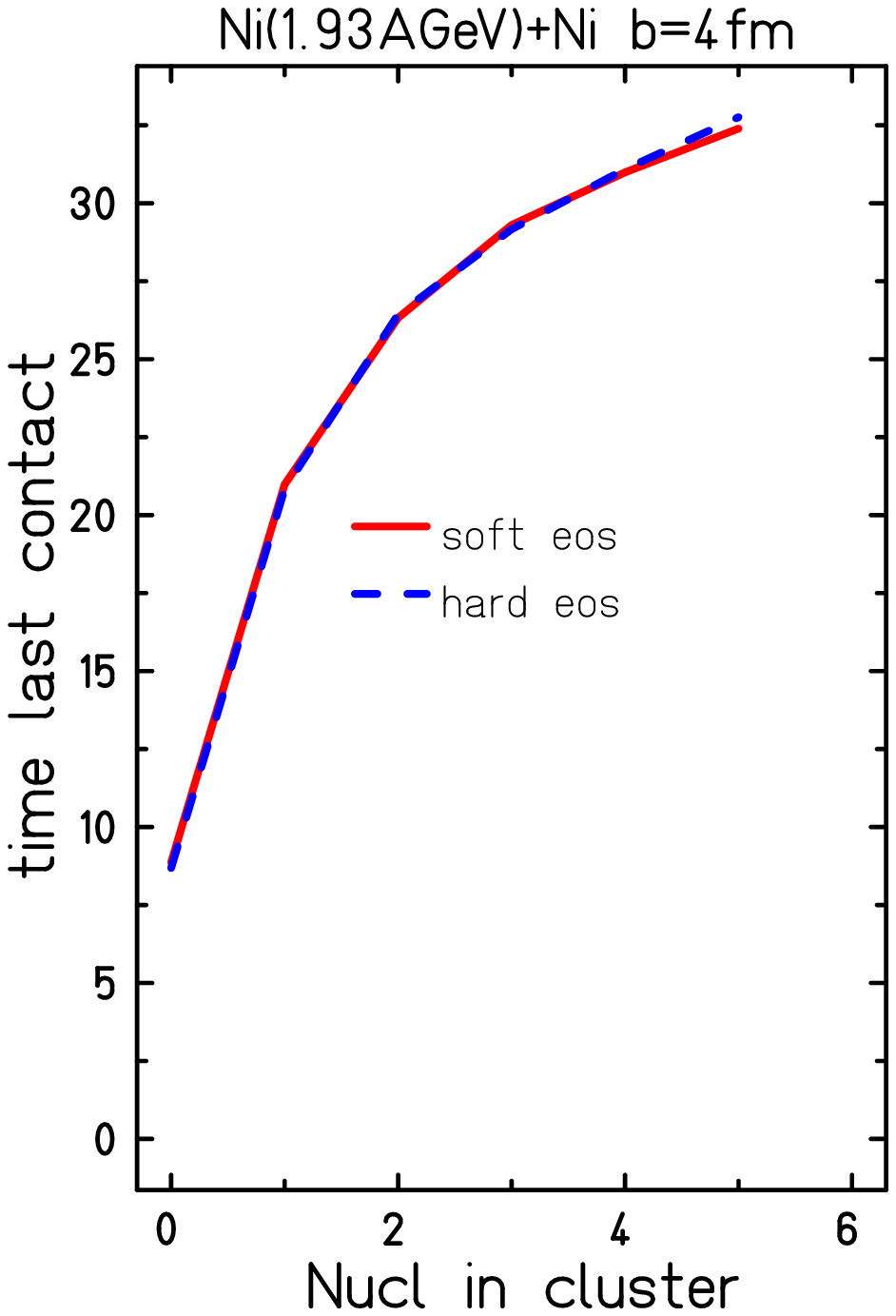}
\caption{Freeze-out density, mean collision number of the hyperon and freeze-out
time of the hyperon as function of the cluster size}
\label{fig:density}
\end{figure}
As we have demonstrated in the previous section, the hyperons have to undergo
rescattering in order integrate a fragment. This effect is even more pronounced
when going to larger hyperclusters.
As already seen in the mid of fig \ref{fig:pxrho} practically all hyperons found in a
hypertriton show a very low freeze-out density. 
This feature remains for other hyperclusters as it can be depicted from the l.h.s. of fig \ref{fig:density}, which presents the density
(in units of the ground state density) at which the last collisional contact between the hyperon and the nuclear matter takes place.
While unclustered hyperons freeze out at a density visibly higher than ground state density, while clustered
hyperons freeze out at about on third of normal matter density.

The mid part of  Fig.~\ref{fig:density} presents the total number of collisions the hyperon has undergone in
the reaction. Even if unclustered hyperons have already undergone almost 3 collisions with nucleons, the number
of collisions increases strongly with the size of the cluster. In hypertritons the hyperons have already
collided more than 7 times in the average. This means that many collisions are necessary in order to arrange
the hyperon that way in nuclear matter that way that they can be bound into
isotopes by the potentials.

The r.h.s. finally gives the "freeze-out time" of the hyperons, i.e the time when the last collision of a hyperon
with a nucleon has happened. To give a reference point, the maximum density is reached after about 4-5 fm/c and
the passing time (time which would be needed by the projectile to pass the
target) is
roughly 9 fm/c. This time corresponds exactly to the mean freeze-out time of unclustered nucleons. However, the
formation of a hypertriton needs about 2-3 times the passing time and larger hyperclusters still stay in
contact for more than 40 fm/c. This demonstrates again that a significant time of "rearrangement" is needed in
order to allow a hyperon to take place in a cluster.

\section{FRIGA results}
\begin{figure}[hbt]
\centering
\includegraphics[width=\linewidth]{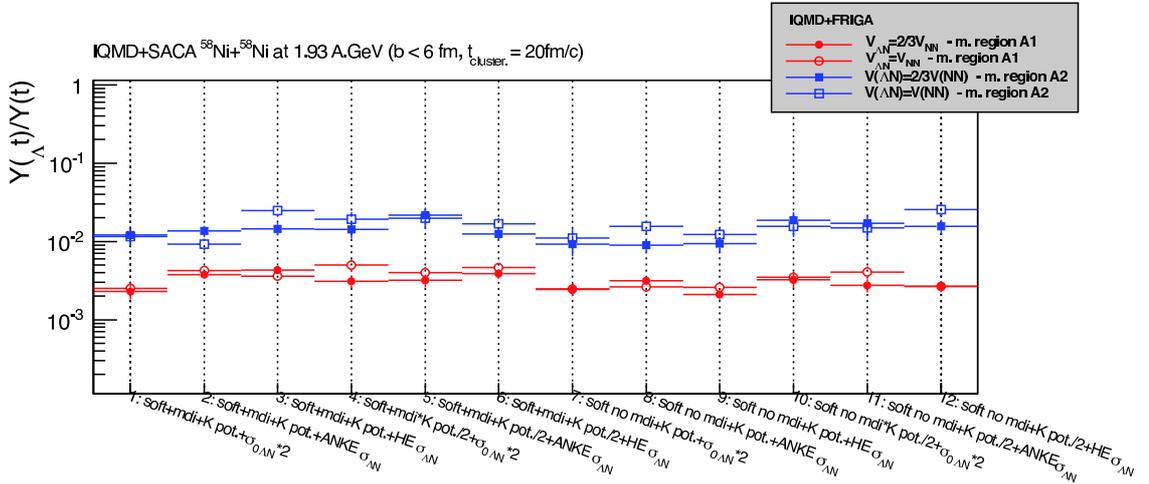}
\caption{FRIGA results on the ratio hypertriton/triton applying FOPI
acceptance cuts}
\label{fig:FRIGA}
\end{figure}
At the end of this article we want to summarize briefly several effects which
can influence the absolute hypertriton yields. For this purpose we present on
the r.h.s. of Fig.~\ref{fig:FRIGA} the results of the FRIGA model on the ratio of
hypertritons to tritons obtained in two different phase space regions accesible
to FOPI. These phase space regions correspond to the rapidity marked by the
triangles in Fig.~\ref{fig:ratio}. It should be note that experiment detects the
hypernuclei by analysis of the vertex of the decay products $^3$He and $\pi^-$
(which have to be detected in a sufficiently large phase space region), by
dtermination of the invariant mass and by subtraction of combinatorical 
background. These constraints diminish the accessible region to two rather small 
areas.  

FRIGA is a novel fragmentation approach described more in detail in
\cite{arnaud}. It is working similarly to the SACA model
\cite{gos97} applying a Metropolis algorithm in order to find the configuration
of maximum binding energy of the clusters. Like SACA it includes Skyrme type
interactions and momentum dependent interactions but also surface and asymetry 
energies. There is the additional possibility to include pairing energy and
shell effects.  In order to treat hypernuclei correctly FRIGA also includes
hyperon-nucleons interactions, which are normally assumed to correspond to two third of
the Skyrme potentials of nucleons.  These results are presented in Fig.~\ref{fig:FRIGA}  
by full symbols. Additionally calculations
assuming the hyperon-nucleon interactions to be identical to the nucleon Skyrme 
interactions are presented by open symbols. We see that the second option
changes the hypertriton yield slightly, since the binding of hyperons in nuclear
matter is changed with the forces.

We see additionally the influence of the rescattering cross sections and the kaon optical potential. These results confirm
the previous indications about the influence of these quantities: 
enhancing the cross sections enhances the hypertriton yields. Diminuishing the
kaon optical potential enhances also the hyperon yield and thus the number of
hypertritons again.
The preselection of parameters was restrained to calculations using a soft
equation of state since the analysis of kaon data clearly indicates that the
experimental data can only be explained by the use of a soft EOS \cite{fuchs,kaoneos}.
Preliminary results of the FOPI collaboration on this ratio exist, but still 
under reanalysis, thus an official release has not been published yet.

\section{Conclusion}
In conclusion we have demonstrated that the formation of hypertritons is strongly
affected by the hyperon-nucleon rescattering, which allows the hyperons to enter
the phase space of the clusters remaining from the spectator remnants.
For that purpose a high number of rescattering is necessary.
Hypernuclei show thus a low freeze-out density and a late freeze-out time.  
Their kinematical properties are strongly aligned to the behaviour of the
spectator matter.
Further analysis has to be done in order to allow for detailed comparison to
nuclear data.


\begin{thebibliography}{99}
\bibitem{schaffner} J. Schaffner, C.B. Dover, A. Gal, C. Greiner and H.
St\"ocker, Phys. Rev. Lett. 71, 1328 (1993)
\bibitem{pano} P. Papazoglou, S. Chramm, J. Schaffner-Bielic, H. St\"ocker and
W. Greiner, Phys. Rev. C57, 2576 (1998)
\bibitem{hashimoto} O. Hashimoto and H. Tamura, Prog. Part. Nucl. Phys. 57, 564
(2006)
\bibitem{star} Y-G Ma (Star collaboration), EPJ Web Conf 66, 04020 (2014)
\bibitem{saito} T.R. Saito et al (HypHI collaboration) Nucl. Phys. A881 (2012)
218
\bibitem{alice} R. Lea, (Alice collaboration), Nucl. Phys. A914, 415 (2013)
\bibitem{rap13}Ch. Rappolt et al. Nucl. Phys. A913 (2013) 170
\bibitem{gaitanos} Th. Gaitanos, H. Lenske, U. Mosel, Phys. Lett B 675 (2009) 297
\bibitem{steinheimer} J. Steinheimer, K. Gudima, A. Botvina, I. Mishustin, M.
Bleicher, H. St\"ocker, Phys. Lett. B 714 (2012) 85
\bibitem{botvina} A. Botvina, J. Steinheimer, E. Bratkovskaja, M. Bleicher, J.
Pochodzalla, Phys. Lett. B 742 (2015) 7
\bibitem{arnaud} A. Le F\`evre, Y.Leifels, J. Aichelin, Ch. Hartnack, V. Kireyev
and E. Bratkovskaya, arXiv:1509.06648v1, to be published in J. Phys. G
\bibitem{rap15}  Ch. Rappold et al., Phys. Lett. B747 (2015) 129. 
\bibitem{IQMD}
C.~Hartnack, R.~K.~Puri, J.~Aichelin, J.~Konopka, S.~A.~Bass, H.~Stoecker and W.~Greiner
Eur.\ Phys.\ J.\  A {\bf 1}, 151 (1998)

\bibitem{Hartnack-PR}
  C.~Hartnack, H.~Oeschler, Y.~Leifels, E.~L.~Bratkovskaya and J.~Aichelin,
  Phys.\ Rept.\  {\bf 510} (2012) 119
\bibitem{ANKE}
  M.~B\"uscher {\it et al.},
  Phys.\ Rev.\  C {\bf 65}, 014603 (2002)
  [arXiv:nucl-ex/0107011].

\bibitem{gos97} P.B. Gossiaux, R. Puri, Ch. Hartnack, J. Aichelin, Nucl. Phys. A \textbf{619} (1997) 379-390.

\bibitem{fuchs}
  C.~Fuchs, A.~Faessler, E.~Zabrodin and Y.~M.~Zheng,
  Phys.\ Rev.\ Lett.\  {\bf 86}, 1974 (2001)
  [arXiv:nucl-th/0011102].

\bibitem{kaoneos}
  C.~Hartnack, H.~Oeschler and J.~Aichelin,
  Phys.\ Rev.\ Lett.\  {\bf 96}, 012302 (2006)
  [arXiv:nucl-th/0506087].


\end{thebibliography}
\end{document}